# NEW CLASS OF PSEUDORANDOM D-SEQUENCES TO GENERATE CRYPTOGRAPHIC KEYS

B. Prashanth Reddy
Oklahoma State University, Stillwater

**Abstract**

This article proposes the use of pseudorandom decimal sequences that have gone through an additional random mapping for the design of cryptographic keys. These sequences are generated by starting with inverse prime expansions in base 3 and then replacing 2 in the expansion with either the string 01 or 10 based on the preceding bit, which represents a general mapping. We show that the resulting pseudorandom sequences have excellent autocorrelation properties. Such a method can be extended to inverse prime expansions to any base.

**I. Introduction**

The effort of finding new families of pseudorandom sequences is a part of the process of keeping one step ahead of eavesdroppers and they are important in the design of cryptographic keys that cannot be easily guessed by intruders [1]-[5]. It is in this spirit we propose the use of a new class of pseudorandom decimal sequences [6]-[8] that are generated by starting with inverse expansions in radix 3 and then replacing 2 in the expansion with either the string 01 or 10 based on the preceding bit and the proposed idea can be extended to sequences in any base. These random number generators are efficient, although by their very nature they must be periodic. Their use is ubiquitous in information systems [9]-[14]. Decimal sequences of different kind have been discussed in the literature [15],[16].

D-sequences have finite periodic length which is produced by the decimal expansion of the inverse prime number q in a modulo r division. The sequence $a_i$ is generated according to the rule:

$$a_i = [\, r^i \bmod q \,] \bmod r \qquad (1)$$

where q is a prime number. These sequences are strings of bits generated in a very convenient and efficient manner. Our goal is to perform another mapping

$$b_i = f(a_i | a_{i-1}) \qquad (2)$$

where the output binary bit $b_i$ depends on the previous output value. The function $f(.)$ can take a variety of forms depending upon our objectives. Here we consider an elegant function where it chooses from one of two substrings based on the previous digit produced by the mapping. We would like to do this in a manner that increases the cryptographic strength of the $b$-string.

As is well-known, inverse prime expansions (D- or decimal sequences) possess good autocorrelation properties and so they are excellent candidates for pseudorandom sequences. But they have an obvious structural redundancy that is a disadvantage: for a maximum length decimal sequence, the bit sequence in the first half of the period is the complement of the second half of the period. This paper addresses this limitation by generating binary sequences by starting with ternary sequences and then replacing the 2s by 0s and 1s in a manner that the frequencies of the bits are not changed. Such a process can be applied to decimal expansions in any base and, therefore, the ternary representation should be seen only as an example of the method.

**II. Unbalanced ternary D-sequences**

If the base chosen for the D-sequence is r = 3, then the sequence created will be a ternary sequence of 2s, 1s and 0s. Ternary pseudo-random D-sequences are generated using a formula similar to (1) as below:

$$a_i = [\ 3^i \bmod q\ ] \bmod 3 \tag{3}$$

where q is a prime number. The maximum length (q-1) sequences are generated when 3 is a primitive root of q.

Autocorrelation for a normalized ternary pseudo-random decimal sequences is calculated by using the below formula (3).

$$c(k) = \frac{1}{n} \sum_{j=0}^{n} a_j \cdot a_{j+k} \tag{4}$$

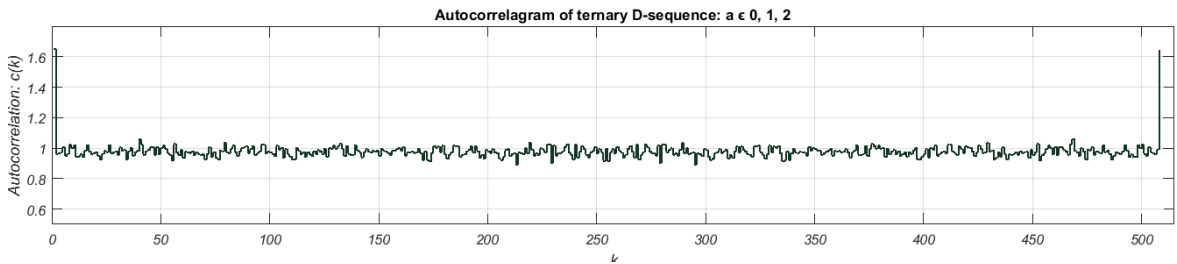

Figure 1: Unbalanced ternary D-sequence autocorrelation fir prime number 509

Figure 1 presents the autocorrelation function of an unbalanced ternary pseudo-random D-sequence considering a prime number 509. As we know from decimal sequence theory since the period is 508, each of the three digits 0, 1, and 2 is almost equally likely.

For all values of $a_i$, $a_i \in \{0, 1, 2\}$. Since the digits have equal probability, the peak value of the autocorrelation function is

$$1/3 \sum_{i=0}^{2} i^2 = 1.667$$

It is clear that the D-sequences are not cryptographically strong, but that is not the property that is essential in generating cryptographic keys. In any event, our objective is to strengthen these sequences by means of an additional mapping (2).

### III. Balanced ternary D-sequences

Due to the computational convenience, balanced ternary sequences (where the 2s have been replaced by -1s) are preferred when compared to unbalanced ternary sequences. Also, we can find better autocorrelation properties when the $a_i$ sequence normalized to stream of bits containing 0s, 1s and -1s (where 2s are represented as -1s).

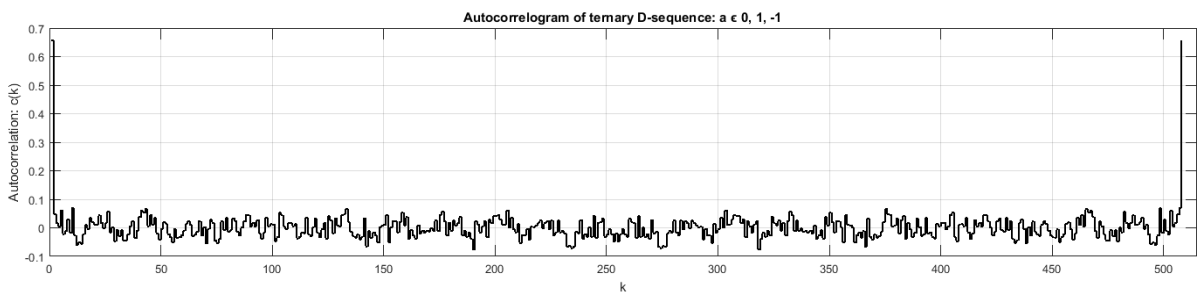

Figure 2: Balanced ternary D-sequence correlogram considering a prime number 509

Although the autocorrelation function looks better its peak has been reduced to 0.6 due to $1/3 \sum_{i=-1}^{1} i^2 = 0.667$. To overcome this limitation we propose to replace the 2s of the ternary sequence by 01 or 10 based on the preceding bit.

## III. Unbalanced ternary D-sequences

Here in this approach, we will replace encountered 2 in $a_i$ bit stream with '01' if it's preceded value in the bit stream is 0, or '10' if its preceded value in bit stream is 1. When substrings of several 2s are encountered, then the above scheme of mapping can be used recursively as shown in Figure 3.

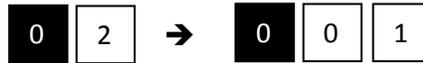

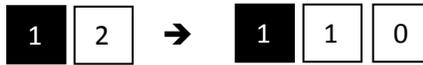

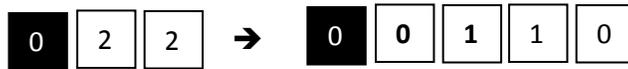

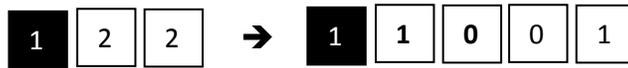

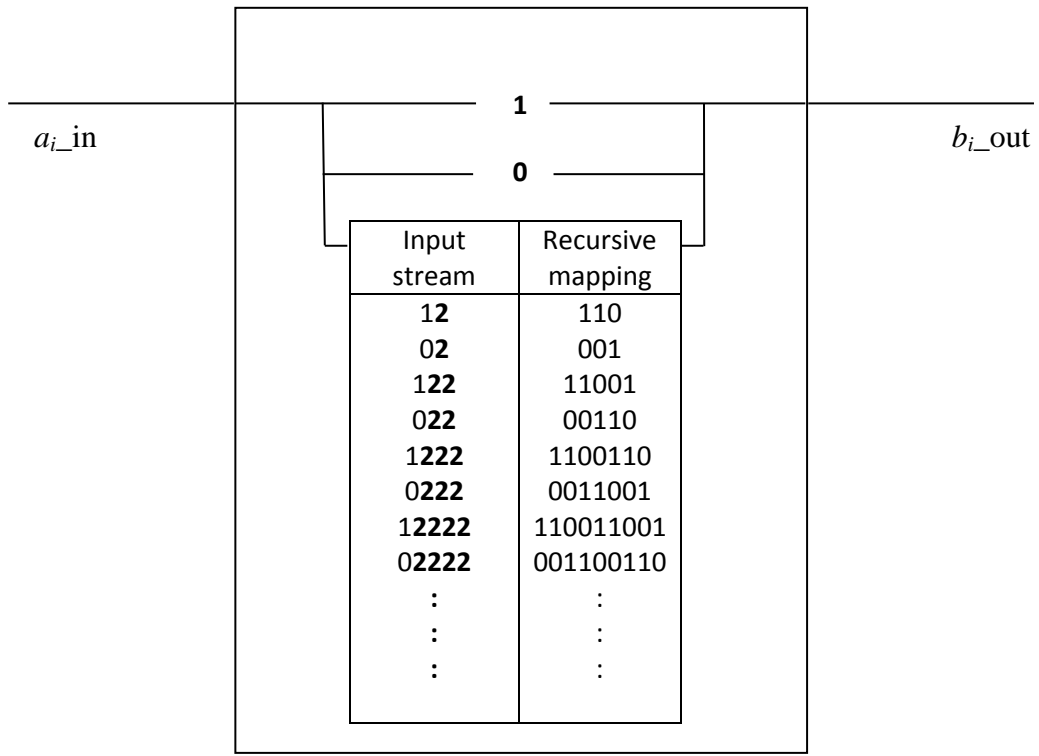

Figure 3: Bit stream mapping

By this approach, one can generate a variable sized bit stream for $a_i$ by challenging adversaries to guess the periodic length of the underlying ternary D-sequence.

*Illustration:* For a prime number 7, $a_i$ = [0 2 0 1 2 1] with a periodic length of 6 and its normalized ternary sequence will be $b_i$ = [0 0 1 0 1 1 0 1] with a periodic length of 8. Likewise, the *a*-sequence [1 0 2 2 1 1 0] will be transformed to the b-sequence [1 0 0 1 2 1 1 0] which in the next step to [1 0 0 1 1 0 1 1 0].

Figure 4, presents the autocorrelation properties for this ternary pseudo-random D-sequence with the peak autocorrelation of 0.5 due to the unbalanced nature of the sequence.

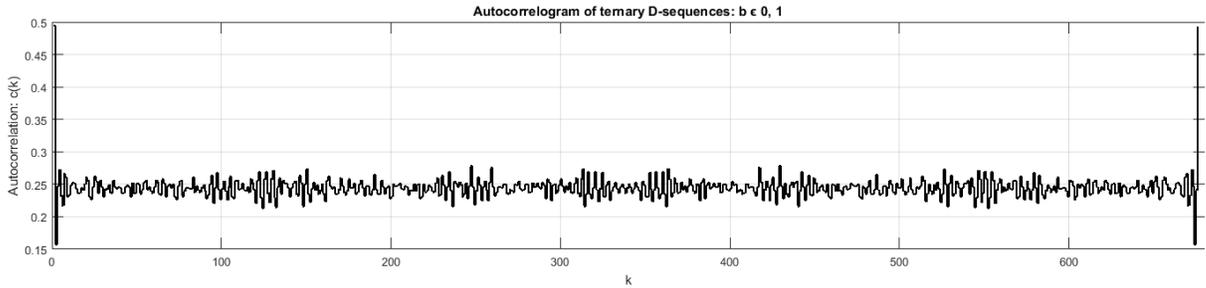

Figure 4: Autocorrelation for the b-sequence

Generating a predictable pseudorandom sequence such as a decimal sequence, or exposing even a few bits of random sequence in each of several digital signatures, may suffice to obtain the private key since such sequences are not cryptographically strong. Clearly the additional nonlinearity introduced in the sequence will make the task cryptographically harder.

**IV. Balanced b-sequences**

The method of the use of ternary b-sequences will provide no opportunity for adversaries to estimate the periodic length of D-sequences. Variable periodic length is advantageous in dealing with adversaries.

Table 1: Calculated lengths of enhanced ternary D-sequences for different prime numbers

| Variable lengths of PN ternary D-sequence depending occurrence of 2 in $b_i$ | | |
|---|---|---|
| Prime | Total number of 2's in $a_i$ | Length of enhanced ternary D-sequence: $b_i$ |
| 509 | 168 | 676 |
| 593 | 194 | 786 |
| 599 | 190 | 788 |
| 643 | 226 | 868 |
| 719 | 199 | 917 |
| 769 | 199 | 967 |
| 797 | 232 | 1028 |
| 827 | 228 | 1054 |
| 883 | 236 | 1118 |
| 907 | 221 | 1127 |
| 991 | 236 | 1226 |
| 1021 | 221 | 1241 |
| 1171 | 236 | 1406 |

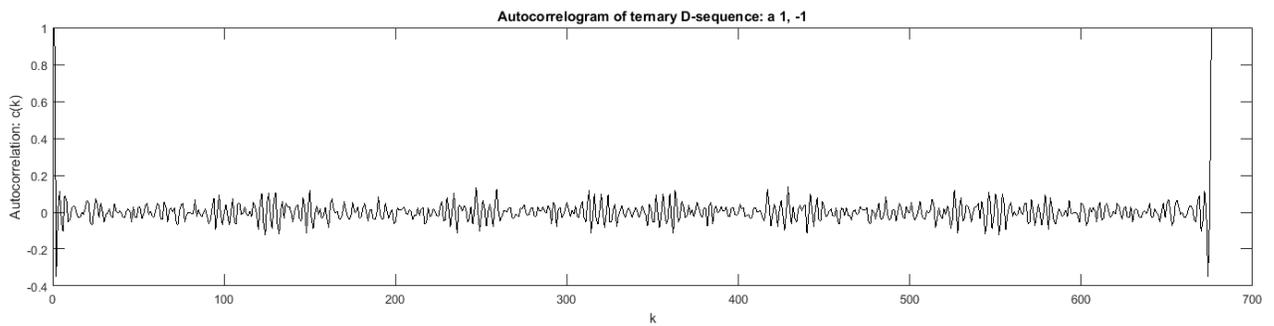

Figure 5: Autocorrelation for balanced new class of b-sequence

The $b_i$ sequence is also represented in the balanced form where 0s have been mapped into -1s. Figure 5 presents the autocorrelation properties for a new class of ternary pseudo-random D-sequence when the recursive bit stream is balanced with better autocorrelation.

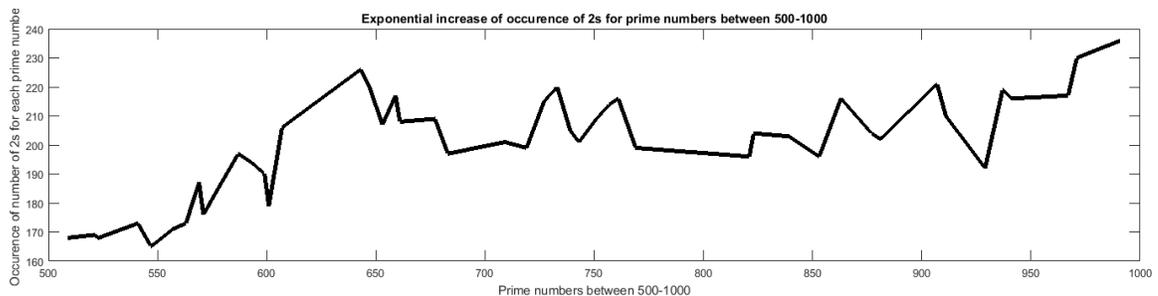

Figure 6: Increase of 2s in $a_i$ for each prime number between 500-1000

The increase is not systematic since the patterns of consecutive 2s will be encountered in an unexpected way and, furthermore, not all sequences are maximum length as the primes are increased. This is at the basis of the increased cryptographic strength of the b-sequences.

## V. Conclusions

We proposed the use of ternary pseudo-random D-sequences, together with a mapping to take the 2s into 0s and 1s that leads to a new class of binary random sequences that can be good candidates for cryptographic keys.

One of the major advantages of the proposed class of sequences is the flexibility in the choice of the period. This method of pseudorandom sequence generation need not be limited to ternary expansions and can be used for expansion to any radix with a subsequent mapping to binary.

## VI. REFERENCES

[1] T. Ritter, Randomness tests: a literature survey, http://www.ciphersbyritter.com/RES/RANDTEST.HTM, 2002.
[2] R. Landauer, The physical nature of information. Physics Letters A 217: 188-193, 1996
[3] S. Kak, The initialization problem in quantum computing. Foundations of Physics 29: 267-279, 1999.
[4] S. Kak, Quantum information and entropy. Int. Journal of Theo. Phys. 46: 860-876, 2007.
[5] A. Kolmogorov, Three approaches to the quantitative definition of information, Problems of Information Transmission, 1, 1-17, 1965.
[6] S. Kak and A. Chatterjee, On decimal sequences. IEEE Transactions on Information Theory, IT-27: 647 – 652, 1981.

[7] S. Kak, Encryption and error-correction coding using D sequences. IEEE Transactions on Computers, C-34: 803-809, 1985.

[8] J. Bellamy, Randomness of D sequences via diehard testing. arXiv:1312.3618

[9] A. J. Paul. C and A. Scott. Handbook of Applied Cryptography, Library of Congress Cataloging-in-Publication Data, 1965.

[10] S. Even, O. Goldreich, A. Lempel, A randomized protocol for signing contracts. Comm. of the ACM 28: 637-647 (1985)

[11] L. Washbourne, A survey of P2P Network security. arXiv:1504.01358, 2015.

[12] S. Kak, The Architecture of Knowledge. CSC, New Delhi, 2004.

[13] F. Rocha, S. Abreau, M. Correia, The final frontier: confidentiality and privacy in the cloud. IEEE Computer, vol. 44, September 2011.

[14] J. Garay, R. Gennaro, C. Jutla, T. Rabin, T. Secure distributed storage and retrieval. Theoretical Computer Science, volume 243, issue 1-2, pages 363 – 389, 2000.

[15] S.R.K. Gangasani, Testing D-sequences for their randomness. arXiv:0710.3779, 2007.

[16] S.B. Thippireddy, Binary random sequences obtained from decimal sequences. arXiv:0809.0676, 2008.